# On the Utility of Virtual Staining for Downstream Applications as it relates to Task Network Capacity

Sourya Sengupta[1,2], Jianquan Xu[4], Phuong Nguyen[2,3], Frank J. Brooks[2], Yang Liu[2,3,4], and Mark A. Anastasio*[1,2,3]

[1]Department of Electrical and Computer Engineering, University of Illinois Urbana-Champaign, Urbana, IL, USA

[2]Center for Label-free Imaging and Multiscale Biophotonics, University of Illinois Urbana-Champaign, Urbana, IL, USA

[3]Department of Bioengineering, University of Illinois Urbana-Champaign, Urbana, IL, USA

[4]Departments of Medicine and Bioengineering, University of Pittsburgh, Pittsburgh, PA, USA

* Corresponding Author

**Email:** maa@illinois.edu

**Abstract**

Virtual staining, or in-silico-labeling, has been proposed to computationally generate synthetic fluorescence images from label-free images by use of deep learning-based image-to-image translation networks. In most reported studies, virtually stained images have been assessed only using traditional image quality measures such as structural similarity or signal-to-noise ratio. However, in biomedical imaging, images are typically acquired to facilitate an image-based inference, which we refer to as a downstream biological or clinical task. This study systematically investigates the utility of virtual staining for facilitating clinically relevant downstream tasks (like segmentation or classification) with consideration of the capacity of the deep neural networks employed to perform the tasks. Comprehensive empirical evaluations were conducted using biological datasets, assessing task performance by use of label-free, virtually stained, and ground truth fluorescence images. The results demonstrated that the utility of virtual staining is largely dependent on the ability of the segmentation or classification task network to extract meaningful task-relevant information, which is related to the concept of network capacity. Examples are provided in which virtual staining does not improve, or even degrades, segmentation or classification performance when the capacity of the associated task network is sufficiently large. The results demonstrate that task network capacity should be considered when deciding whether to perform virtual staining.

1. Introduction

Simultaneously tracking the dynamics and interactions of different organelles and subcellular structures within a cell is critical for elucidating cellular functions[1–4]. Fluorescence microscopy has been instrumental in providing highly specific visualization through intrinsic or extrinsic markers, but it is constrained by phototoxicity, photobleaching, and limited fluorophore multiplexing capacity, restricting its utility for high spatial and temporal resolution imaging of multiple organelles. As an alternative, label-free imaging leverages intrinsic optical or chemical properties to generate contrast without exogenous labels, employing modalities like phase contrast, differential interference contrast (DIC), autofluorescence, and quantitative phase imaging (QPI)[5–7]. The limited molecular specificity of label-free methods, such as QPI, has motivated the development of virtual staining techniques. These techniques typically employ deep learning-based image-to-image translation (ITIT) models to computationally predict synthetic fluorescence from label-free images[8–11]. Although various terms such as "digital staining", "stainless staining", and "in-silico labeling" have been used interchangeably in the literature, this work will henceforth use the term "virtual staining."

Most studies of virtual staining assessed the quality of virtually stained images using pixel-based image quality (IQ) metrics such as Pearson correlation coefficient (PCC), structural similarity index (SSIM), peak signal-to-noise ratio (PSNR), and mean squared error (MSE)[3,4,12–14]. However, in biomedical imaging, images are typically acquired to perform downstream clinical or biological tasks, and traditional IQ measures, like SSIM, PSNR, or MSE, may not indicate the utility of an image for such purposes[15]. Recent studies have demonstrated the use of virtually stained images for tasks that include segmentation, tracking, and counting nuclei to analyze cellular dynamics and growth, as well as for high-throughput phenotypic screening to assess cell states using deep learning models[10,11,14,16–19].

Previous studies of image denoising and super-resolution demonstrated that the impact of ITIT methods on the performance of downstream tasks is not universally beneficial[15,20]. Specifically, when deep neural networks were used to perform downstream tasks, which will be referred to as task networks, it was demonstrated that the effectiveness of ITIT was dependent on the capacity of the task networks. Here, a machine learning network's capacity refers to its ability to approximate complex functions or represent relationships between inputs and outputs. In effect, it defines how much information a network can process and generalize from the training data[21].

A direct way to understand task network capacity is to consider the number of trainable parameters it contains; networks with more parameters ideally should be capable of extracting more useful information to perform a task than a network with fewer parameters. However, when training data are limited in number, the capacity of a network is not completely utilized. In these cases, specific regularization techniques like early stopping — where training is halted when performance on validation data stops improving for a few epochs — are routinely used[22] to prevent overfitting by effectively limiting network capacity. In these cases, the notion of learnable capacity is relevant[23], which relates to the portion of the network's capacity that can be realized given the available (and limited) training data and training process. For a given network, a specific training process, and a regularization strategy, the learnable capacity increases with increasing training dataset size[23]. Unless needed to distinguish them, henceforth, capacity and learnable capacity will both be referred to simply as capacity in this article.

Segmentation and classification remain pertinent tasks in virtual staining pipelines in bioimaging applications. However, to the best of our knowledge, the reported studies on virtual staining have not systematically investigated how task network capacity influences the utility of virtually stained images with consideration of these tasks. The primary goal of this study is to accomplish this in studies that involve label-free cellular images. Segmentation and classification task performance is computed when task networks of varying capacity are utilized with label-free, virtually stained, and fluorescence images, and the results are compared and analyzed.

2. Results

2.1. Workflow

A dataset of label-free quantitative phase images of cancer cells—both untreated and drug-treated was assembled, with each image precisely registered to its corresponding fluorescence counterpart at the pixel level. Co-registered label-free and fluorescence images were collected using the omni-mesoscope imaging system[24], a versatile and cost-effective multimodal imaging platform. By including both untreated and drug-treated cells, the dataset captures a diverse range of nuclear morphologies associated with biological processes such as mitosis, proliferation, apoptosis, and stress response.

Virtual staining was performed to translate label-free images to corresponding fluorescence images. In the case studies described below, two different downstream tasks were considered: cell nuclei segmentation and classification of cell state after drug treatment. Deep neural networks were employed to perform these tasks, referred to as task networks, using 1) label-free quantitative phase images; 2) virtually stained images; and 3) the directly acquired fluorescence images. The performance of different task networks of varying capacities was assessed. The goal of this study design was to demonstrate how the capacity of a task network can influence the utility of virtually stained images.

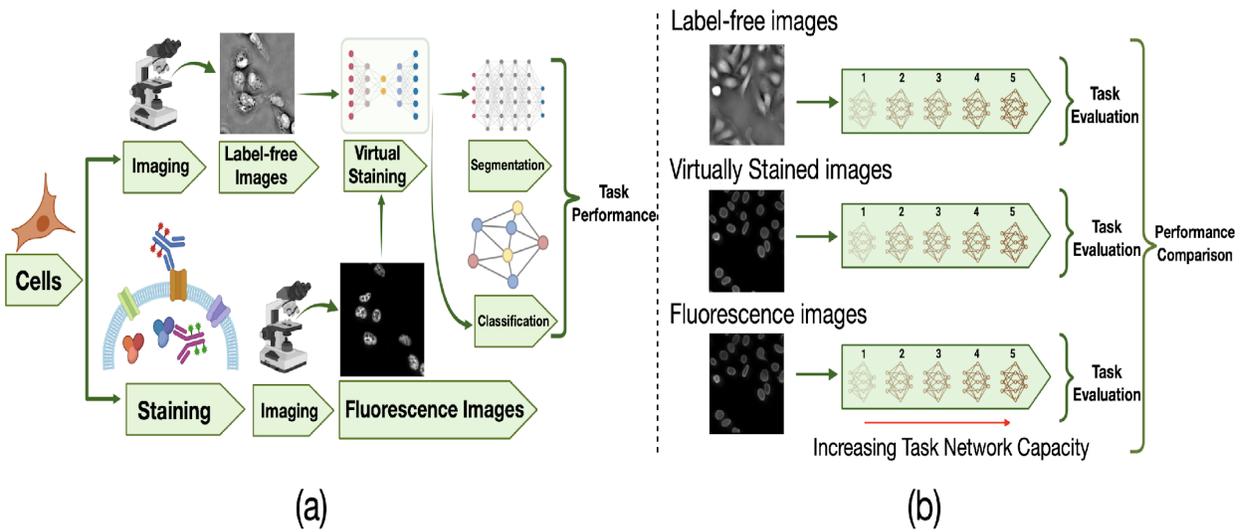

Figure 1: Overview of the study workflow and task-performance evaluation strategy. **(a)** General schematic of the virtual staining framework. These virtually stained images were generated using ITIT. **(b)** Schematic of the task-performance evaluation strategy with varying task network capacity. The virtually stained images, along with label-free and ground-truth fluorescence counterparts, were independently evaluated by use of task networks of varying capacity.

2.1. Case study 1: Binary cell nuclei segmentation task

Cell nuclei segmentation is a fundamental task in biological image analysis as it enables the precise identification and quantification of individual cell dynamics. This is critical for assessing cell cycle states, analyzing spatial relationships in heterogeneous populations. In this study, the impact of virtual staining on the binary cell nuclei segmentation performance was evaluated. The capacity of the segmentation network was systematically varied to assess how it impacts the utility of virtual staining for the segmentation task. Performance comparisons were conducted that employed label-free quantitative phase images, virtually stained images, and ground-truth fluorescence images.

2.1.1. Dataset details

As the fluorescence nuclei marker, Lamin A/C was used. Lamin A/C is a nuclear membrane protein that specifically highlights the nuclear boundary, serving as a reliable marker for cell nuclei detection. Binary segmentation ground truth masks were generated using the open-source Cellpose segmentation pre-trained model[25] directly from the fluorescence images. From the stitched images with a size of 19,000 × 19,000 pixels, random crops of 256 × 256 pixels were extracted to construct the training dataset. The full image consisted of around 15000 cells. A total of 16,000 patch pairs were generated for training, with an additional 1,000 pairs for validation and 3,000 pairs for testing. Corresponding binary mask patches were also cropped from the Cellpose-generated segmentation maps.

2.1.2. Virtual staining

The ITIT models were trained to generate virtually stained Lamin A/C marker images from input label-free patches. Two state-of-the-art ITIT models, Pix2pixHD[26] and EfficientUNet7[27], were employed for this task. Figure 2 shows examples of label-free input image patches, virtually stained images with both the Pix2pixHD and EfficientUNet7 models. Virtually stained images with Pix2pixHD achieved SSIM and PSNR of 0.84 and 25.8, respectively, while EfficientUNet7 achieved SSIM and PSNR values of 0.87 and 26.1. These values represent averages computed over the test set data.

The virtual staining outputs achieved high PSNR and SSIM values, demonstrating their similarity to the ground truth fluorescence images. However, the bounding boxes in Fig. 2 highlight two examples of some notable errors in generated images. In the top row, the red bounding boxes identify a cell undergoing mitosis. In this scenario, the intensity variations in the original label-free image result in a bright glare in the virtual staining output. This glare is less pronounced in the ground truth fluorescence images. This demonstrates how intensity differences in the input label-free image can influence the virtual staining prediction. In the bottom row, the green bounding box indicates an area where two distinct cells, clearly separated in the ground truth fluorescence image, appear merged in the prediction.

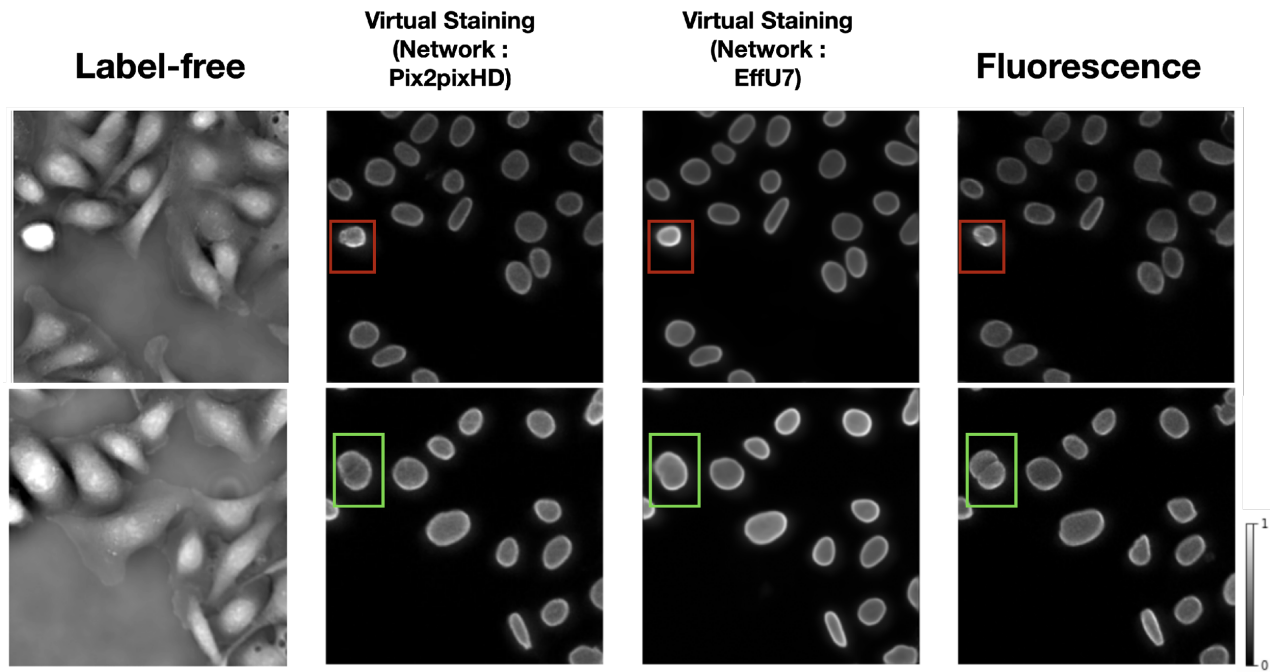

Figure 2: From left: Input label-free image patch, Pix2pixHD-based virtual staining output, EffU7-based virtual staining output, ground truth fluorescence marker images. Examples of errors in virtual staining predictions compared to ground truth fluorescence images are shown in bounding boxes. The red bounding boxes show intensity differences in a mitotic cell, leading to a bright glare in the virtual staining output. The green boxes highlight a case where closely spaced cells are incorrectly merged in the prediction.

2.2.3. Impact of virtual staining on segmentation performance

The effectiveness of virtual staining for improving performance on the binary nuclei segmentation task was assessed with consideration of segmentation networks of varying capacities and learnable capacities. The task network capacity was varied by changing the depth and number of trainable parameters in the segmentation networks. Three progressively deeper U-Net models (Unet1, Unet2, Unet3) were evaluated alongside EfficientUNet0 and EfficientUNet7 models, resulting in a total of five segmentation networks with increasing capacity. Segmentation performance was quantitatively evaluated using the Dice score, precision, and recall. In each case, 15000 image-segmentation mask pairs were utilized for training. Mean precision, recall, and Dice scores were computed from all the images of the test set.

In a subsequent study, the network architecture was fixed at EfficientUNet7, and the network's learnable capacity was increased by increasing the size of the training dataset. The training process and the regularization strategy remained same for all the cases. Datasets comprising 100, 300, 1,000, 5,000, 10,000, and 15,000 image-segmentation mask pairs were used for training. For each dataset size, the training images were randomly partitioned into three independent subsets of the same size. The mean precision, recall, and Dice score were computed on each of these splits, and the mean and standard deviation across these three splits were reported to quantify the task network's performance.

The results of both studies are shown in Figs. 3 and 4. In the case of limited capacity (in terms of trainable parameters) or a limited learnable capacity (in terms of training images), virtual staining networks helped to improve segmentation performance compared to that achieved by the use of the label-free images. However, if the segmentation network has sufficient capacity, the task network trained with original label-free images performed similarly to the virtually stained label-free images.

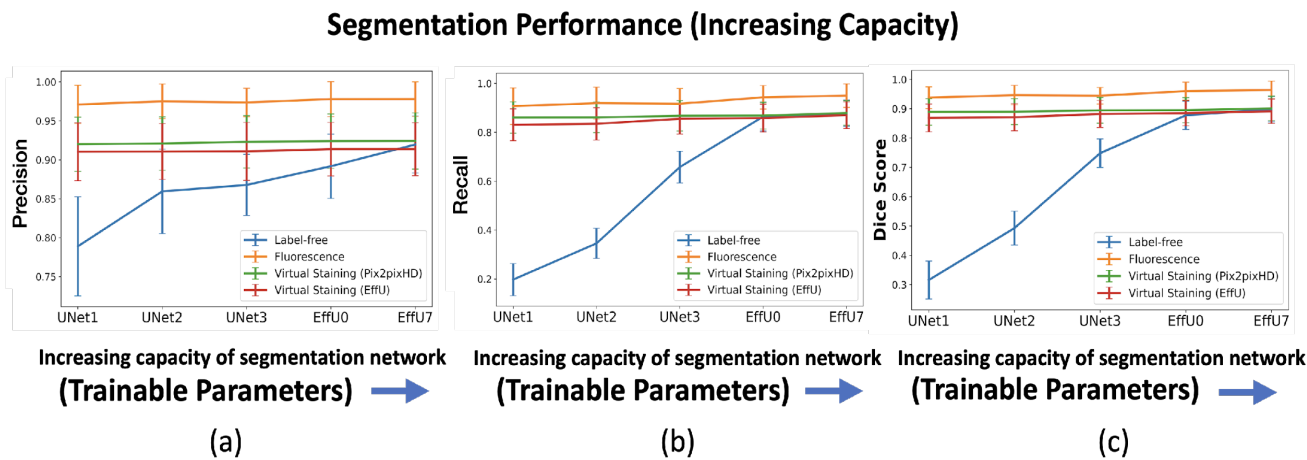

Figure 3: Segmentation performance vs. network capacity (number of training images). **(a)** precision, **(b)** recall, **(c)** Dice scores using label-free, fluorescence, virtually stained images generated by the EffUNet, Pix2pixHD. In all cases, the results show that virtual staining can improve the performance when the capacity is low, but for a higher capacity segmentation network virtual staining does not improve the performance compared to the network trained with the original label-free images

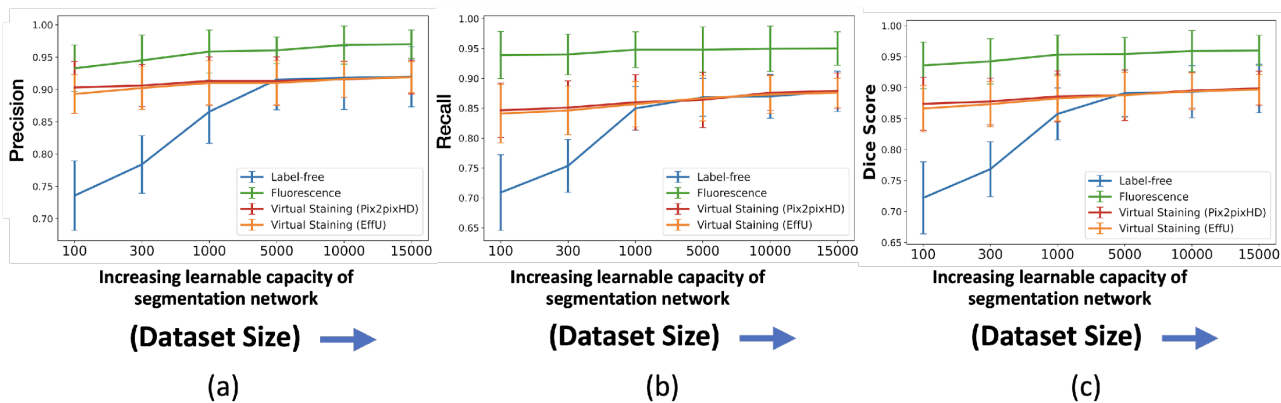

Figure 4: Segmentation performance vs. network learnable capacity (number of training images). **(a)** precision, **(b)** recall, **(c)** Dice scores using label-free, fluorescence, virtually stained images generated by the EffUNet, Pix2pixHD. In all the cases, the results show that virtual staining can improve the performance when the number of training images is low, but when the number of training images is increased virtual staining does not improve the performance compared to original label-free images.

To gain insights into why, in certain cases, virtual staining helped but not in others, the predicted nuclei segmentation masks are visualized in the Fig. 5. The results show the predicted masks for label-free, virtually stained, and fluorescence images using segmentation networks of increasing capacity. It can be observed that for label-free images, the lower-capacity networks segmented high-contrast nuclei, which were typically in mitotic phases (red bounding box), while other nuclei in interphase were poorly segmented. However, as the capacity of the segmentation network increased, the performance on label-free images improved, and the segmentation predictions became comparable to those obtained using virtually stained images. This trend was also reflected in Fig. 3, where low-capacity networks demonstrated poor performance on label-free images, while higher-capacity networks achieved higher performance similar to that obtained with virtually stained images. This indicates that virtual staining acted effectively as an advanced pre-processing step, simplifying the segmentation task for lower-capacity networks. However, for segmentation networks that have higher capacity, the advantage yielded by pre-processing diminished, resulting in minimal or no additional performance gains. It is also noteworthy that the task performance achieved using the virtually stained outputs was lower than that obtained using the corresponding ground truth fluorescence images, potentially due to the loss of fine boundary information or cell merging artifacts previously discussed in Sec. 2.2.2.

The masks predicted with segmentation networks of increasing learnable capacity are visualized in Fig. 6. Like Fig. 5, Fig. 6 reveals that, for lower learnable capacity network (lower training data in this case), virtual staining enhanced segmentation performance as compared to the use of the label-free images. However, this advantage diminished with increasing learnable capacity. This is consistent with the performance trend shown in Fig. 4.

To better understand how different segmentation networks processed label-free and stained images, the intermediate feature maps were examined. Feature maps provide insight into what the network focuses on at different stages of learning. By analyzing the penultimate decoder layer of each segmentation network, it can be observed how the network refines its attention to relevant structures, such as cell nuclei in this case, as its capacity increases.

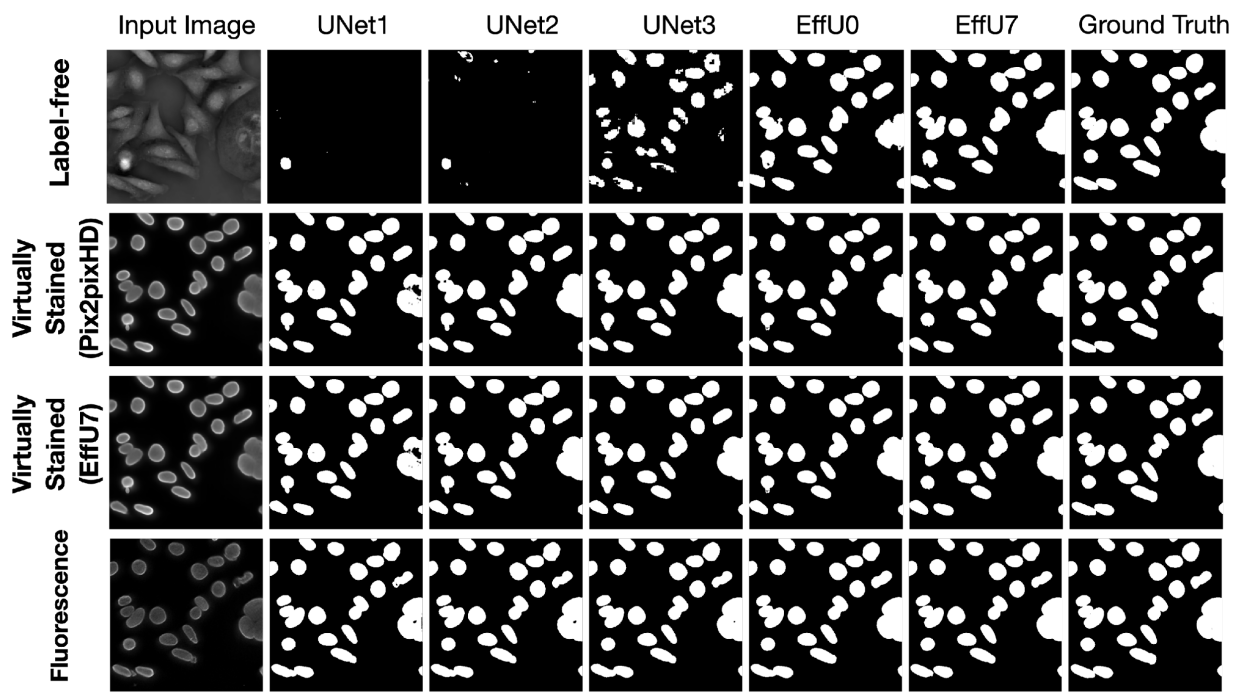

Figure 5: The top row displays segmentation predictions for label-free images, the middle row for virtually stained images, and the bottom row for fluorescence images. The columns from left to right show the input image, segmentation predictions from networks of increasing capacity (Unet1 to EffU7), and the ground truth mask (last column). The results show that as network capacity increases, the segmentation performance of label-free images becomes comparable to that of virtually stained images.

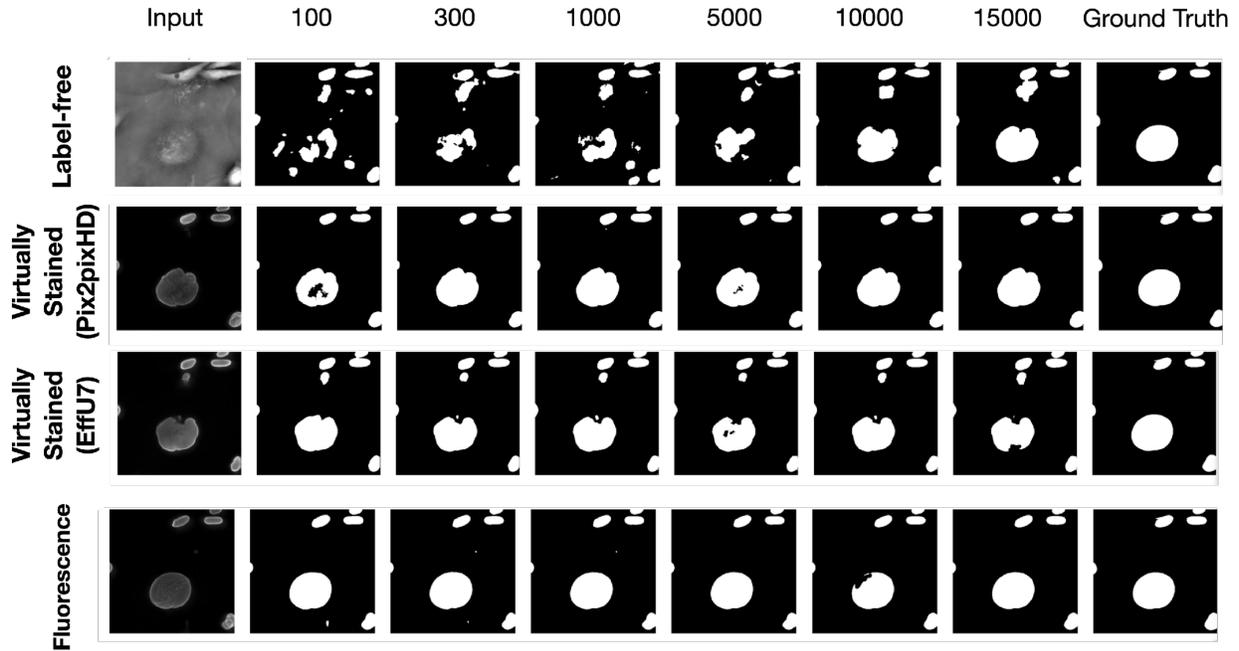

Figure 6: The top row displays segmentation predictions for label-free images, the middle row for virtually stained images, and the bottom row for fluorescence images. Columns from left to right show the input image, segmentation predictions from networks of increasing learnable capacity (training data size 100 to 15000), and the ground truth mask (last column). The results show that as network learnable capacity increases, the segmentation performance of label-free images becomes comparable to that of virtually stained images.

Figure 7 presents the mean feature maps of the penultimate decoder layer for three segmentation networks of varying capacities: Unet2 (low capacity), EffU0 (medium capacity), and EffU7 (high capacity). As shown by the red bounding box, for the low-capacity Unet2 network, the feature maps for label-free images show a broader focus, highlighting entire cells rather than specifically targeting the nuclei. In contrast, the same network produces feature maps for virtually stained and fluorescence images that concentrate on cell nuclei boundaries, demonstrating more targeted segmentation. As the capacity of the network increases, the feature maps for label-free images progressively shift to resemble those of virtually stained images, with both focusing exclusively on the cell nuclei. These results also indicate that when the task network has a high capacity, the post-processing (e.g., virtual staining) does not provide additional benefit in learning the region of interest

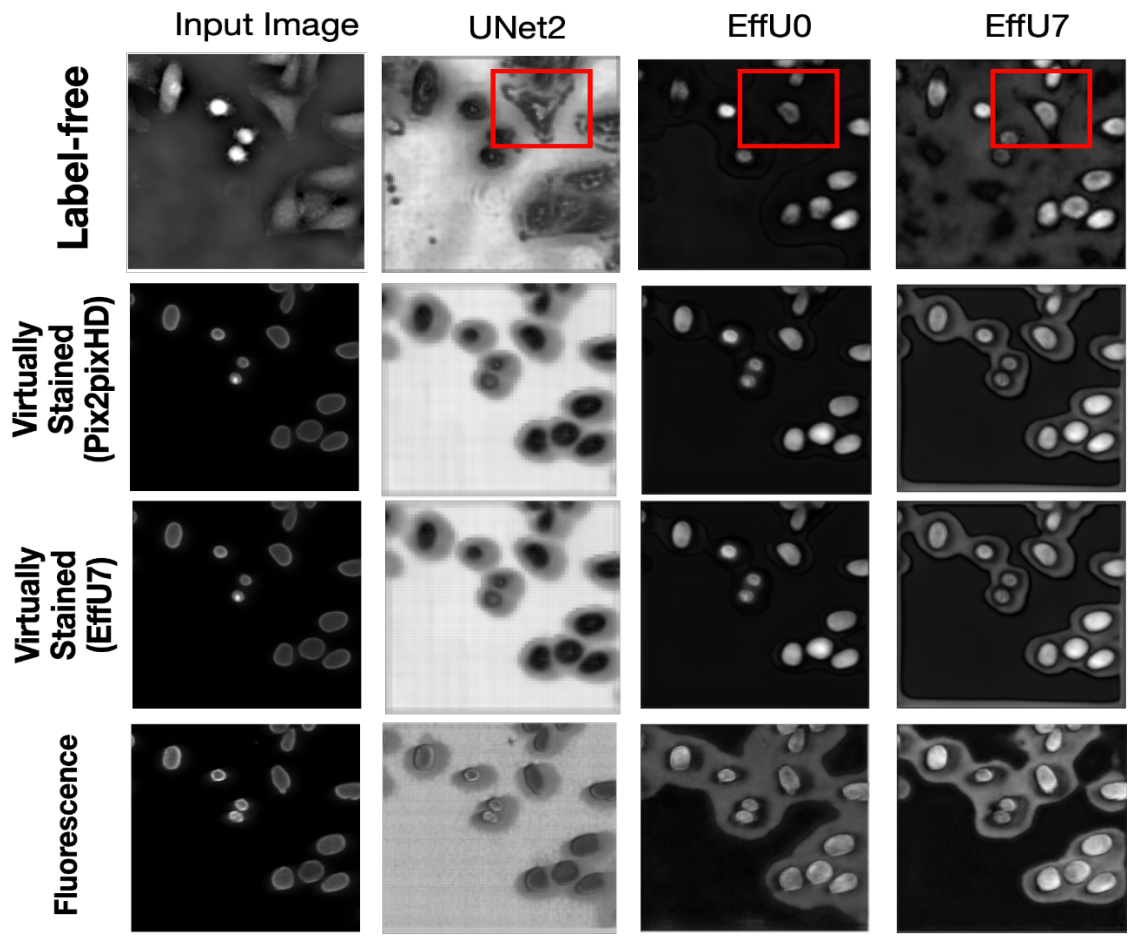

Figure 7: Columns (from left to right) display the input image followed by the mean feature maps of the penultimate decoder layer for three segmentation architectures of increasing capacity—Unet2 (low), EfficientUNet0 (medium), and EfficientUNet7 (high). Rows (top to bottom) correspond to the four image modalities provided to the networks: (1) label-free images, (2) virtually stained images with Pix2pixHD, (3) virtually stained images with EfficientUNet7, and (4) ground-truth fluorescence images. As shown by the red bound boxes, with the increasing capacity of the segmentation network, the mean decoder map focused more on cell nuclei instead of the complete cells for label-free images, like the virtually stained maps. But for the lower capacity network, the label-free images could not focus on the cell nuclei.

2.3. Case study 2: Cell state classification after drug treatment

Accurately classifying the temporal stage of a cell line post-drug treatment is crucial for understanding the dynamics of drug response and resistance mechanisms. This case study investigates the usefulness of virtual staining for the task of classifying control vs drug-treated cells (20 days after treatment) by use of cropped cell nuclei. The classification network's capacity or learnable capacity was varied systematically to understand the impact of virtual staining on classification performance. A comparison was performed with label-free, virtually stained, and ground truth fluorescence images.

2.3.1. Dataset details

SW480 cancer cells were treated with the 5-FU drug. The images were captured before drug treatment (control class) and 20 days after drug treatment (treated class) using the omni-mesoscope imaging system. Quantitative phase images of single-cropped cell nuclei and H3K27me3 fluorescence marker images of single-cropped cell nuclei were employed in this study. The H3K27me3 fluorescence marker, which highlights the trimethylation of lysine 27 on histone H3, provides critical insights into chromatin state and epigenetic regulation within nuclei[28]. This marker is closely associated with transcriptionally repressive regions, enabling the identification of silenced genomic domains that influence cell behavior. A total of 10000 pairs (label-free, fluorescence pair) of cropped cell nuclei were used for training the virtual staining network with an additional 1000 pairs for validation and 1000 pairs for testing. In each case, the dataset was divided between control and treated classes in a 1:1 ratio.

2.3.2. Virtual staining

For the classification task, virtual staining was performed using single-cell nuclei. Both Pix2pixHD and EfficientUNet models were employed for virtual staining. Figure 8 shows examples of input label-free images, virtual stained outputs produced by the Pix2pixHD and EfficientUNet7 networks, and ground truth fluorescence cell nuclei. The virtually stained images produced by the use of the Pix2pixHD achieved SSIM and PSNR values of 0.965 and 29, respectively. The virtually stained images yielded by the EfficientUNet7 network achieved SSIM and PSNR values of 0.97 and 29.1, respectively. In both cases, these values represent a mean over all test images. While both the Pix2pixHD and EffU7 networks yielded similar PSNR and SSIM values, the EffU7 results were found to be much smoother compared to Pix2pixHD outputs in terms of texture from a qualitative subjective assessment.

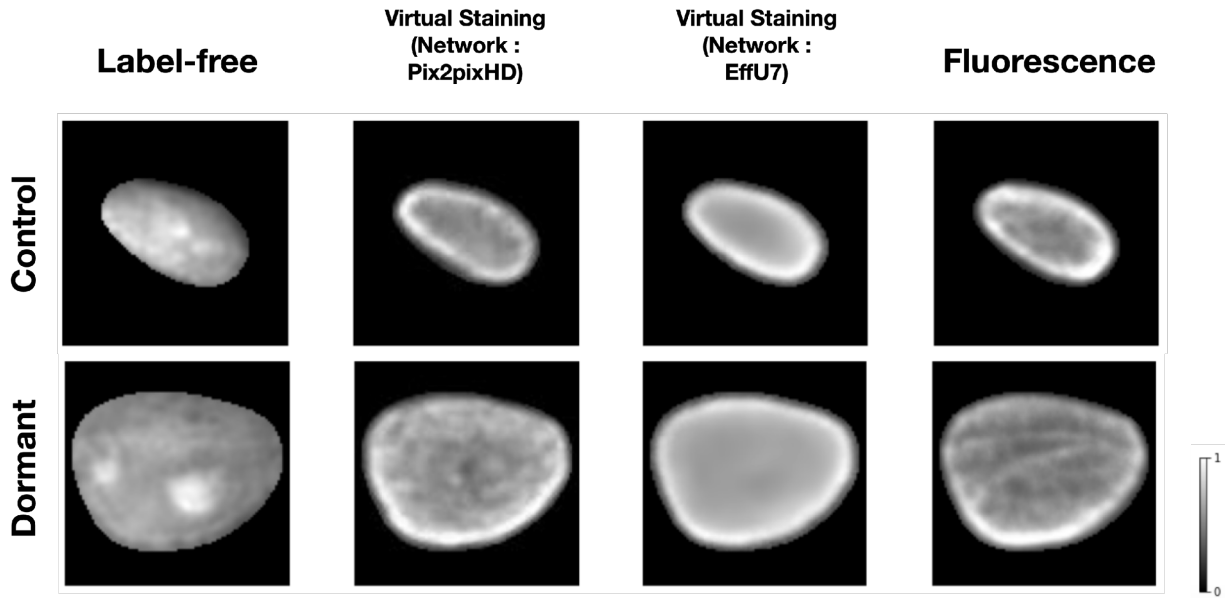

Figure 8: From left: Input label-free cell nuclei, Pix2pixHD-based virtual staining output, EffU7-based virtual staining output, ground truth fluorescence marker cell nuclei. The top row corresponds to one example of the control class, the bottom row corresponds to one example of the treated class (20 days after treatment). Both the virtual staining networks achieved high SSIM and PSNR values for the outputs.

2.3.3. Impact of virtual staining on classification performance

Like the segmentation task, the influence of virtual staining on the classification task was investigated by systematically varying the capacity and learnable capacity of the task networks a.k.a. deep learning-based classifiers. Two shallow convolutional neural networks (CNN1 and CNN3) were used along with deeper architectures, including ResNet18, ResNet50, and ResNet101, resulting in a total of five classifiers with increasing capacity. CNN1 consisted of a single convolutional layer followed by ReLU activation, max-pooling, a fully connected dense layer, and a sigmoid activation for classification. CNN3 included three convolutional layers, each followed by ReLU activation and max-pooling, and ended with a dense layer and sigmoid activation. Classification performance was measured using the area under the receiver operating characteristics (ROC) curve (AUC).

In a subsequent study, the classifier network architecture was fixed at ResNet101, and the learnable capacity was increased by increasing the size of the training dataset. The regularization strategy of early-stopping was fixed for all the cases. Datasets comprising 100, 500, 1,000, 5,000, 8,000, and 10,000 images were used for training. For both studies, the ROC software[29] developed by Metz and colleagues was used to compute the AUC values and associated uncertainty estimates. The values were computed by fitting the ROC curve using a binomial model by the software[29].

The results of the two studies are shown in Figs. 9 and 10. In each case, the left subfigure presents the AUC values produced using task networks of increasing capacity, trained on label-free, fluorescence, and virtually stained images generated by the Pix2pixHD network. The right subfigure shows the corresponding AUC values produced using task networks of increasing capacity, trained on label-free, fluorescence, and virtually stained images generated by the EffUNet network. Both show that, in the case of a limited capacity classification network, virtual staining enhanced classification performance. However, when the classification network possessed increased capacity, virtual staining degraded the classification performance compared to that computed directly by use of the label-free images

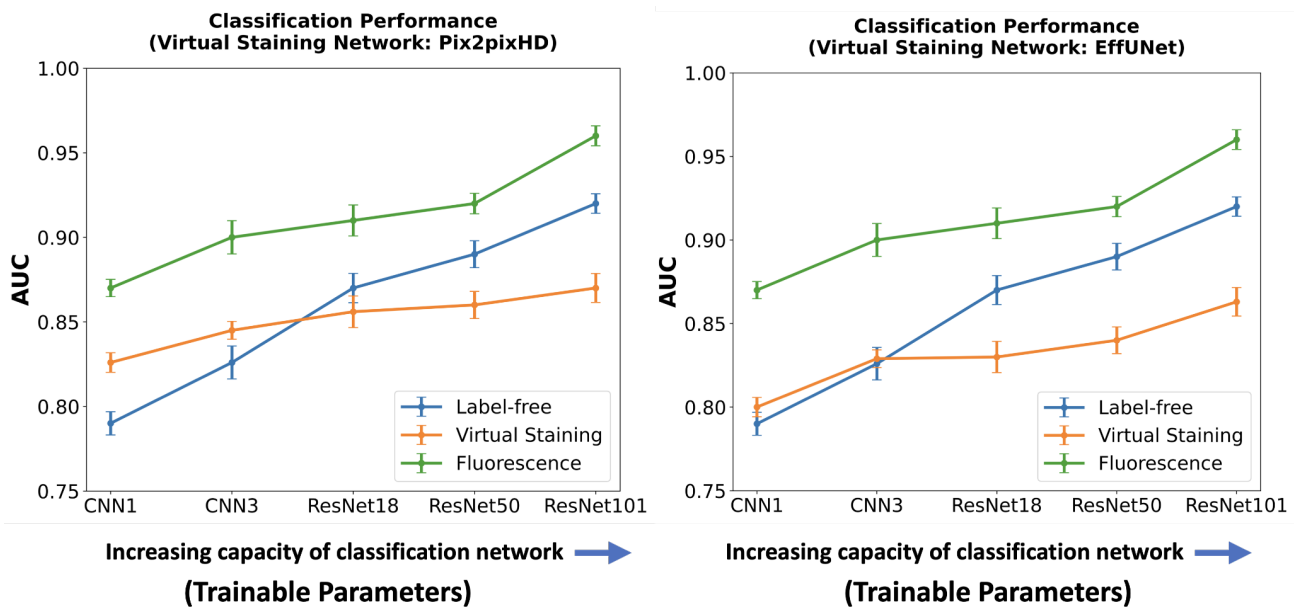

Figure 9: AUC comparison of classification networks with varying capacities for label-free, fluorescence, and virtually stained images. **Left**: AUC performance using virtually stained images generated by the Pix2pixHD network along with label-free and fluorescence. **Right**: AUC performance using virtually stained images generated by the EffUNet7 network along with label-free and fluorescence. The results show that, for the lower capacity classifier, virtual staining gave a performance boost but with the increasing capacity of the classification network, original label-free images yielded higher AUC than virtually stained images.

A similar trend was observed when considering the learnable capacity of the classification network, as shown in Fig. 10. The classification performance degradation is potentially due to the loss of texture information in the virtually stained images.

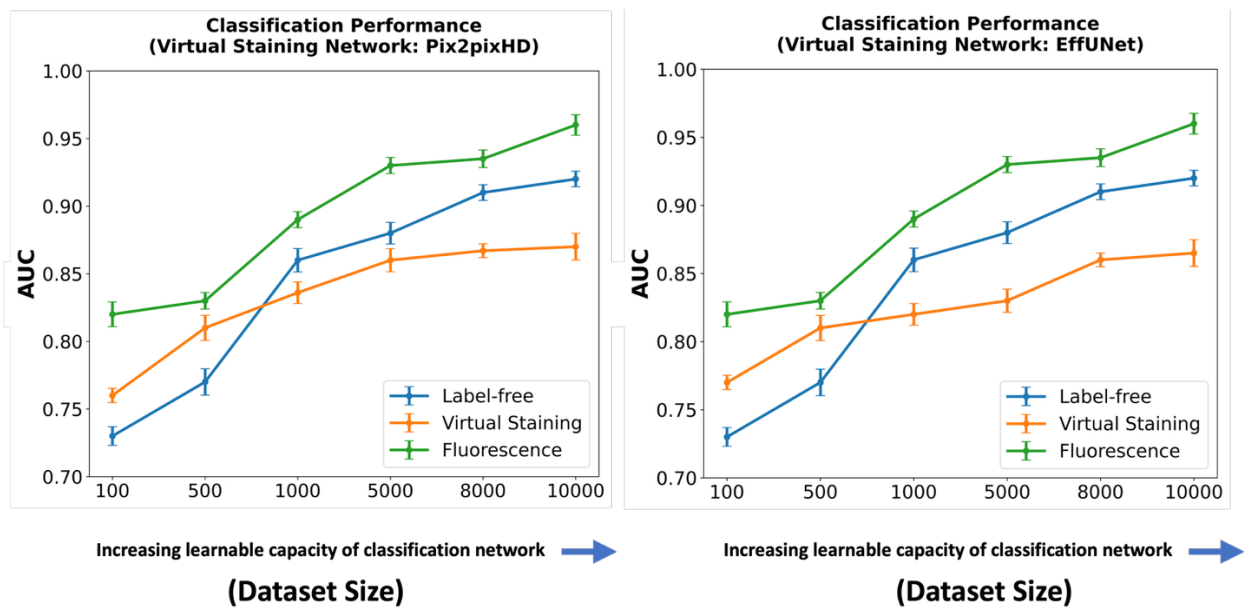

Figure 10: AUC comparison of classification networks with increasing learnable capacity (increasing numbers of training images) for label-free, fluorescence, and virtually stained images. **Left**: AUC performance using virtually stained images generated by the Pix2pixHD network along with label-free and fluorescence. **Right**: AUC performance using virtually stained images generated by the EffUNet7 network along with label-free and fluorescence. The results show that, for a lower number of training images, virtual staining enhances classification performance. However, with the increasing number of training images, this advantage was lost, and classification performance using the original label-free images was higher.

To provide further insights into the above classification results, the low-dimensional feature embeddings of the control class and the treated class were analyzed by comparing their cosine distances. Three classifiers with varying capacities, defined by the number of trainable parameters, were selected: CNN1 (low capacity), ResNet18 (medium capacity), and ResNet101 (high capacity). The low-dimensional embeddings of the feature extraction layers were computed for each of those. Feature embeddings for both classes were extracted from the trained classifiers, and the cosine distances between embeddings were computed. Cosine distance plots for label-free, virtual staining, and ground truth fluorescence images are shown in Fig. 11.

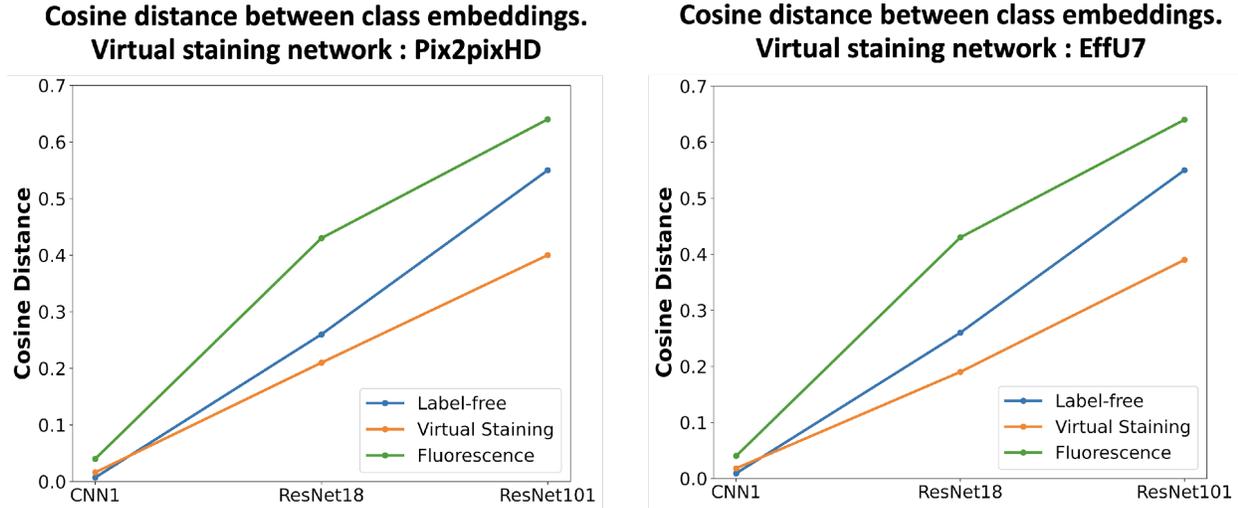

Figure 11: Cosine distances between class embeddings for control and treated classes for different classifier capacities. **Left**: Distances computed for the networks trained with virtually stained images generated by the Pix2pixHD network, label-free, and fluorescence. **Right**: Distances computed using virtually stained images from the EffUNet7 network, label-free and fluorescence. Low-capacity classifiers trained with virtually stained images produce slightly higher cosine distances compared to the network trained with label-free images, whereas medium- and high-capacity classifiers trained with label-free images exhibit greater class separability.

The results indicate that, for the lower capacity classifier (CNN1), the cosine distance was higher for virtually stained images compared to that of the label-free images. However, for medium and high-capacity classifiers (ResNet18 and ResNet101), the original label-free images showed greater cosine distances between the two classes compared to that of with the virtually stained images. The larger cosine distance indicates higher class separability, suggesting an easier classification task. Virtual staining was found to enhance the performance of low-capacity classifiers by reducing the embedding distance between classes, thereby aiding classification. In contrast, for higher-capacity classifiers, virtual staining was observed to degrade performance.

3. Discussion

Recent studies have shown that virtual staining can produce images that possess high traditional IQ measures that include SSIM and MSE[30,31]. However, high traditional IQ might not always correlate with something clinically or biologically meaningful[32]. Therefore, it is critical to evaluate the downstream task performance with the virtually stained images. In this context, segmentation and classification tasks are widely used with microscopy images for different clinical and biological applications. Segmentation tasks are typically employed to delineate cellular structures such as nuclei, cytoplasm, or whole cells. These tasks serve as essential steps for quantifying morphological features, analyzing spatial organization, and extracting single-cell metrics. On the other hand, classification tasks focus on assigning categorical labels to individual cells based on their visual or morphological characteristics. In cell imaging, these tasks include classifying cells into different cell cycle phases, identifying cell states, or distinguishing between drug-treated and untreated phenotypes[10,11,14,16–19].

The results of our studies using both segmentation and classification tasks demonstrated that virtual staining was helpful in specific circumstances, particularly when the task network possessed limited capacity. However, when the task network's capacity was sufficiently high, virtual staining provided no improvement and, in some cases, even degraded performance. Importantly, the presented studies emphasize that virtual staining is not universally advantageous and that its role needs to be critically evaluated depending on the task and the capacity of the task solver. The objective assessment of virtual staining methods needs to be performed with consideration of specific downstream tasks and inference models (e.g., deep networks) employed to perform the task.

The results of the presented studies are intuitively consistent with the data processing inequality (DPI) from information theory [33]. The DPI tells us that any transformation or image processing applied to the raw images cannot increase their task-relevant information content as measured by mutual information[34,35]. Consequently, such transformations cannot improve the performance of an ideal Bayesian observer[36,37] who, by definition, implements an optimal decision strategy and uses all available task-relevant information. Although not ideal observers, the task networks with high capacity or high learnable capacity were able to utilize much of the task-relevant information present in label-free images and did not benefit from processing the label-free image data via virtual staining. However, the lower-capacity task networks, behaving as sub-optimal decision makers, lacked the ability to extract substantial task-relevant information from the label-free image data. In such cases, virtual staining served as an effective preprocessing step, transforming the original label-free image into a representation better exploited by the low capacity classifiers to improve task performance[38].

In conclusion, this work offers insights about the interplay between the capacity of a task network that performs an image-based inference and the utility of virtually stained images. The findings emphasize that task network capacity should be taken into account when deciding whether to apply virtual staining, as its effectiveness can vary across task network capacities for different downstream clinical or biological applications. In practice, virtual staining can be beneficial when training data are limited but offers marginal benefit when large datasets are available to train high-capacity task networks. Although these case studies were conducted using quantitative phase images and two virtual staining networks, Pix2pixHD and EffUNet, the underlying observations are expected to generalize across virtual staining frameworks, as they are grounded in fundamental principles of information processing.

4. Materials and Methods

4.1 Cell culture details and sample preparation

SW480 cells cultured on gelatin-coated coverslips were exposed to 35 $\mu$M 5-FU for 20 days. Live-cell imaging was performed for 5 hours using the omni-mesoscope under physiological conditions (37°C, 5% $CO_2$), after which the cells were fixed in 4% paraformaldehyde for 15 minutes and subsequently stained for fluorescence imaging.

4.2. Imaging system and procedure

The cell images were captured using the omni-mesoscope imaging system[24]. The imaging system had a pixel size of 320 nm and a field of view (FOV) of 5 mm². A transport-of-intensity (TIE) based phase retrieval method was used to obtain quantitative phase images. The SW480 cancer cell line was used for the conducted studies, with cells treated with the chemotherapy drug Fluorouracil (5-FU) to study treatment responses over time. Imaging was performed both on untreated cells (control group) and the treated cells (after treatment with the drug).

4.3. Image pre-processing

The imaging with the omni-mesoscope resulted in multiple patches with an overlap between each. To generate the complete FOV image, a cross-correlation algorithm was used for stitching. Quantitative phase images were registered with fluorescence images using the scale-invariant feature transform (SIFT) feature[39] matching method. All of these steps were performed using the MATLAB Image Processing Toolbox[40]. For the segmentation task, the fluorescence images were directly used to generate binary cell nuclei segmentation masks with the Cellpose[41] pre-trained model. The same masks were also used to crop single-cell nuclei images from the label-free or fluorescence images. The complete image was first multiplied by a mask to isolate specific cell nuclei. Each region of cell nuclei was then cropped and embedded onto a black background of a fixed size of 192 X 192 pixels. A histogram of the cell nuclei images was plotted, and the interquartile range was computed to identify and remove outliers, addressing potential issues such as clustering or segmentation errors. Finally, stray pixels and small spurious regions were removed using a connected component-based filtering approach after each binary segmentation mask was processed by first identifying all connected components via region labeling.

4.4. Training details for virtual staining networks

The training dataset of image patches for the downstream segmentation task comprised 16,000 image patches of size 256 × 256 pixels, with 1,000 images allocated for validation and 3,000 images for testing. Two state-of-the-art virtual staining networks Pix2pixHD[26] and EfficientUNet7[27] were employed. Model evaluation was performed using the weights corresponding to the epoch that yielded the best validation performance. For the classification task, virtual staining was performed at a single-cell level. The training dataset comprised 10,000 pairs of cell nuclei images of size 192 X 192, with 1000 pairs for validation and 1000 for testing. Model evaluation was performed using the weights corresponding to the epoch with the best validation performance. The virtual staining networks were implemented using the Pytorch framework[42] and trained on NVIDIA A100 GPUs.

Pix2pixHD is a conditional generative adversarial network (cGAN) tailored for high-resolution image-to-image translation tasks. It employs a multi-scale generator and discriminator architecture. The generator synthesizes images progressively from coarse to fine resolutions. The Pix2PixHD model was trained using a loss $L_{total}$, which combines adversarial loss and feature matching loss. Mathematically, the loss function can be defined as:

$$L_{\text{total}} = \sum_{k=1}^{K} \mathbb{E}_{X,Y}[\log D_k(X,Y)] + \mathbb{E}_X\left[\log\left(1 - D_k(X, G(X))\right)\right] + \lambda_{\text{FM}} \sum_{k=1}^{K} \mathbb{E}_{X,Y} \sum_{i=1}^{T} \frac{1}{N_i} \left\| D_k^{(i)}(X,Y) - D_k^{(i)}(X, G(X)) \right\|_1 \quad (1)$$

Here, the generator is denoted by $G$, the input image is $X$, the generator output is $G(X)$. The adversarial component of the loss is implemented using a set of multi-scale discriminators $D_k$, each operating at a different image resolution $k \in \{1, ..., K\}$. The term $D_k(X, Y)$ represents the discriminator's output, $D_k(X, G(X))$ corresponds to the output for the generated image pair. The second term in the objective represents the feature matching loss. Specifically, $D_k^{(i)}(\cdot,\cdot)$ denotes the activation of the *i*-th layer of the *k*-th discriminator, and $N_i$ is the number of elements in that layer. The hyperparameter $\lambda_{FM}$ controls the relative weight of the feature matching loss. The value was chosen as 10 following the original implementation[26].

EfficientUNet7 incorporates features from EfficientNet[43], a family of deep learning models optimized for parameter efficiency through neural architecture search. In EfficientUNet7, the encoder employs pre-trained EfficientNet as its backbone, leveraging its depthwise-separable convolutions and squeeze-and-excitation blocks to efficiently capture complex features with fewer trainable parameters. The decoder uses upsampling layers, along with skip connections from the encoder. A hybrid loss function $L_{hybrid}$ that combines mean squared error (MSE) and structural similarity index measure (SSIM) was employed. This hybrid loss can be expressed as:

$$L_{\text{hybrid}} = \alpha \cdot L_{\text{MSE}} + (1 - \alpha) \cdot L_{\text{SSIM}}, \qquad (2)$$

where $L_{MSE} = \frac{1}{N}\sum_{i=1}^{N}(x_i - \hat{x}_i)^2$ is the pixel-wise mean squared error between the target image $x$ and the generated image $\hat{x}_i$, and $L_{SSIM} = 1 - SSIM(x, \hat{x}_i)$ represents the dissimilarity based on the Structural Similarity Index (SSIM). The weighting factor $\alpha$ was selected based on a grid search over values ranging from 0.1 to 0.9, and the best validation performance was consistently observed with $\alpha = 0.9$. The models were trained with a batch size of 16. The Adam optimizer with a learning rate of 3e-4 was used, a stopping criterion was set that the training would stop if there was no decrease in terms of validation loss for 10 consecutive epochs.

4.5. Training details for segmentation networks

Five different segmentation networks were used, UNet1, UNet2, UNet3 [44], EfficientUNet0, EfficientUNet7 [27]. UNet1, the shallowest architecture, was designed with a single convolutional block in the encoder and a single deconvolutional block in the decoder. Each encoder block was composed of a convolutional layer, ReLU activation, and max-pooling operation, while the decoder employed deconvolution, ReLU activation, and concatenation with the corresponding encoder layer. UNet2 was extended to include two blocks per encoder and decoder, while UNet3 included three such blocks. EfficientUNet0 and EfficientUNet7 are two off-the-shelf segmentation networks that employ pre-trained EfficientNet0 and EfficientNet7, respectively, as encoder backbones. The segmentation networks were trained using an Adam optimizer with a learning rate of 1e-4 and a batch size of 4. UNet1, UNet2, and UNet3 were trained from scratch but EfficientUNet0 and EfficientUNet7 were initialized with ImageNet encoder weights. Image patches of 256 X 256 were used to train the model. The segmentation networks were trained using the binary cross-entropy (BCE) loss function[45] $L_{BCE}$. The BCE loss $L_{BCE}$ is defined as:

$$L_{\text{BCE}} = -\frac{1}{N}\sum_{i=1}^{N}\left[y_i \log(p_i) + (1-y_i)\log(1-p_i)\right], \tag{3}$$

where N is the total number of pixels in the image, $y_i \in \{0, 1\}$ is the ground truth label for the i-th pixel, and $p_i \in [0, 1]$ is the predicted probability of the pixel belonging to the foreground (i.e., the positive class). Model performance was quantitatively evaluated using the precision, recall, and Dice score, standard metrics for assessing overlap between predicted and ground truth binary masks. In segmentation, the performance of a predicted mask $\hat{y} \in \{0, 1\}^N$ with respect to a ground truth mask $y \in \{0, 1\}^N$ was evaluated using precision and recall, defined in terms of true positives (TP), false positives (FP), and false negatives (FN) as:

$$\text{Precision} = \frac{TP}{TP + FP} = \frac{\sum_{i=1}^{N} y_i \hat{y}_i}{\sum_{i=1}^{N} \hat{y}_i}, \tag{4}$$

$$\text{Recall} = \frac{TP}{TP + FN} = \frac{\sum_{i=1}^{N} y_i \hat{y}_i}{\sum_{i=1}^{N} y_i}, \tag{5}$$

where $y_i \in \{0,1\}$ is the ground truth label for pixel $i$, $\hat{y}_i \in \{0,1\}$ is the predicted label, and $N$ is the total number of pixels. Using these definitions, the Dice score[46] can be expressed as the harmonic mean of precision and recall:

$$\text{Dice} = \frac{2 \cdot \text{Precision} \cdot \text{Recall}}{\text{Precision} + \text{Recall}}. \tag{6}$$

4.6. Training details for classification networks

Five different classification networks were used, CNN1, CNN3, ResNet18, ResNet50 and ResNet101. The CNN1 and CNN3 were trained from scratch, but ResNet family networks were initialized with ImageNet weights. The classification networks were trained using an Adam optimizer with a learning rate of 1e-4 with a batch size of 16. Single-cell nuclei embedded on a fixed 192 X 192 black background were used for training. A stopping criterion was set that the training would stop if there was no decrease in terms of validation loss for 10 consecutive epochs. All classification networks were trained using the binary cross-entropy (BCE) loss function $L_{BCE}$ defined in Eqn. 3.

## Data Availability

The supporting representative data and results to generate task performances are saved in https://figshare.com/s/d9b8fb91364b12313176.

## Code Availability

All custom code used for training and evaluation will be available at: https://github.com/comp-imaging-sci publicly upon acceptance of the paper.

## Author Information

S.S., J.X., P.N., F.B., Y.L., and M.A.A. are with the University of Illinois Urbana-Champaign, USA.
S.S. is affiliated with the Department of Electrical and Computer Engineering and the Center for Label-free Imaging and Multiscale Biophotonics.
J.X., P.N., and F.B., Y.L. are affiliated with the Center for Label-free Imaging and Multi- scale Biophotonics and the Department of Bioengineering.
M.A.A. is affiliated with all three units.

## Author Contributions

## Acknowledgments

This work was supported in part by NIH Awards P41EB031772, R01EB034249, T32EB019944, R01CA232593 and R01CA254112.

**Conflict of Interests**

The authors declare no competing financial or non-financial interests

**References**

1. Lichtman, J. W. & Conchello, J.-A. Fluorescence microscopy. *Nat. Methods* **2**, 910–919 (2005).
2. Alturkistani, H. A., Tashkandi, F. M. & Mohammedsaleh, Z. M. Histological stains: a literature review and case study. *Glob. J. Health Sci.* **8**, 72 (2015).
3. Ounkomol, C., Seshamani, S., Maleckar, M. M., Collman, F. & Johnson, G. R. Label- free prediction of three-dimensional fluorescence images from transmitted-light microscopy. *Nat. Methods* **15**, 917–920 (2018).
4. Jo, Y. *et al.* Label-free multiplexed microtomography of endogenous subcellular dynamics using generalizable deep learning. *Nat. Cell Biol.* **23**, 1329–1337 (2021).
5. Goswami, N., Anastasio, M. A. & Popescu, G. Quantitative phase imaging techniques for measuring scattering properties of cells and tissues: a review—part i. *J. Biomed. Opt.* **29**, 22713–22713 (2024).
6. Monici, M. *et al.* Cell and tissue autofluorescence research and diagnostic applications. in *Biotechnology Annual Review* vol. 11 227–256 (Elsevier, 2005).
7. Burch, C. & Stock, J. Phase-contrast microscopy. *J. Sci. Instrum.* **19, 71**, (1942).
8. Rivenson, Y. *et al.* Phasestain: the digital staining of label-free quantitative phase microscopy images using deep learning. *Light Sci. Appl.* **8**, 23 (2019).
9. Rivenson, Y. *et al.* Virtual histological staining of unlabelled tissue-autofluorescence images via deep learning. *Nat. Biomed. Eng.* **3**, 466–477 (2019).
10. Ivanov, I. E. *et al.* Mantis: high-throughput 4d imaging and analysis of the molecular and physical architecture of cells. *PNAS Nexus* **3**, 323 (2024).
11. Atwell, S. *et al.* Label-free imaging of 3d pluripotent stem cell differentiation dynamics on chip. *Cell Rep. Methods* **3**, (2023).

12. Christiansen, E. M. *et al.* In silico labeling: predicting fluorescent labels in unlabeled images. *Cell* **173**, 792–803 (2018).

13. Wang, Z., Xie, Y. & Ji, S. Global voxel transformer networks for augmented microscopy. *Nat. Mach. Intell.* **3**, 161–171 (2021).

14. Cheng, S. *et al.* Single-cell cytometry via multiplexed fluorescence prediction by label-free reflectance microscopy. *Sci. Adv.* **7**, 0431 (2021).

15. Li, K., Zhou, W., Li, H. & Anastasio, M. A. Assessing the Impact of Deep Neural Network-Based Image Denoising on Binary Signal Detection Tasks. *IEEE Trans. Med. Imaging* **40**, 2295–2305 (2021).

16. Liu, Z. *et al.* Robust virtual staining of landmark organelles with Cytoland. *Nature Machine Intelligence*, *7*(6), 901-915.

17. Nygate, Y. N. *et al.* Holographic virtual staining of individual biological cells. in *Proceedings of the National Academy of Sciences* vol. 117 9223–9231 (2020).

18. Cross-Zamirski, J. O. *et al.* Label-free prediction of cell painting from brightfield images.

19. Goswami, N. *et al.* Evatom: an optical, label-free, machine learning assisted embryo health assessment tool. *Commun. Biol.* **7**, 268 (2024).

20. Li, K., Zhou, W., Li, H. & Anastasio, M. A. Task-based performance evaluation of deep neural network-based image denoising. in *Medical Imaging 2021: Image Perception, Observer Performance, and Technology Assessment* vol. 11599 114–118.

21. Vapnik, V. N. An overview of statistical learning theory. *IEEE Trans. Neural Netw.* **10**, 988–999 (1999).

22. Prechelt, L. Early stopping-but when? in *Neural Networks: Tricks of the Trade* vols 55–69 (Springer, 2002).

23. Chen, D., Chang, W.-K. & Chaudhari, P. Learning capacity: A measure of the effective dimensionality of a model. (2023).

24. Ma, H., Chen, M., Xu, J., Zhao, Y. & Liu, Y. An omni-mesoscope for multiscale high-throughput quantitative phase imaging of cellular dynamics and high-content molecular characterization. (2024).

25. Stringer, C. & Pachitariu, M. Cellpose3: one-click image restoration for improved cellular segmentation. (2024).

26. Wang, T.-C. *et al.* High-resolution image synthesis and semantic manipulation with conditional gans. in *Proceedings of the IEEE Conference on Computer Vision and Pattern Recognition* 8798–8807 (2018).

27. Baheti, B., Innani, S., Gajre, S. & Talbar, S. Eff-unet: A novel architecture for semantic segmentation in unstructured environment. in *Proceedings of the IEEE/CVF Conference on Computer Vision and Pattern Recognition Workshops* 358–359 (2020).

28. Lanzuolo, C., Lo Sardo, F., Diamantini, A. & Orlando, V. Pcg complexes set the stage for epigenetic inheritance of gene silencing in early s phase before replication. *PLoS Genet.* **7**, 1002370 (2011).

29. Metz, C. E. *et al.* Basic principles of roc analysis. in *Seminars in Nuclear Medicine* vol. 8 283–298 (Elsevier, 1978).

30. Oh, J.-H., Falahkheirkhah, K. & Bhargava, R. Detecting hallucinations in virtual histology with neural precursors. (2024).

31. Huang, L. *et al.* Autonomous quality and hallucination assessment for virtual tissue staining and digital pathology. (2024).

32. Elmalam, N., Nedava, L. B. & Zaritsky, A. In silico labeling in cell biology: Potential and limitations. *Curr. Opin. Cell Biol.* **89**, 102378 (2024).

33. Beaudry, N. J. & Renner, R. An intuitive proof of the data processing inequality. (2011).

34. Shannon, C. E. A mathematical theory of communication. *Bell Syst. Tech. J.* **27**, 379–423 (1948).

35. Ashok, A., Baheti, P. K. & Neifeld, M. A. Task-specific information: an imaging system analysis tool. in *Visual Information Processing XVI* vol. 6575 122–129.

36. Kersten, D. & Mamassian, P. Ideal observer theory. *Encycl. Neurosci.* **5**, 89–95 (2009).

37. Zhou, W., Li, H. & Anastasio, M. A. Approximating the ideal observer and hotelling observer for binary signal detection tasks by use of supervised learning methods. *IEEE Trans. Med. Imaging* **38**, 2456–2468 (2019).

38. Slonim, N. The information bottleneck: Theory and applications. (Hebrew University of, Jerusalem Jerusalem, Israel, 2002).

39. Lindeberg, T. Scale invariant feature transform. *Scholarpedia* **7**, 10491 (2012).

40. Morris, T. Image processing with matlab. Supporting Material for COMP27112 14.

41. Stringer, C., Wang, T., Michaelos, M. & Pachitariu, M. Cellpose: a generalist algorithm for cellular segmentation. *Nat. Methods* **18**, 100–106 (2021).

42. Imambi, S., Prakash, K. B. & Kanagachidambaresan, G. R. Pytorch. in *Prakash*.

43. Tan, M. & Le, Q. Efficientnet: Rethinking model scaling for convolutional neural networks. in *International Conference on Machine Learning* 6105–6114.

44. Ronneberger, O., Fischer, P. & Brox, T. U-net: Convolutional networks for biomedical image segmentation. in *Medical Image Computing and Computer-assisted Intervention–MICCAI 2015: 18th international conference* vol. proceedings, part III 18 234–241 (Springer, Munich, Germany, 2015).

45. LeCun, Y., Bengio, Y. & Hinton, G. Deep learning. *Nature* **521**, 436–444 (2015).

46. Zou, K. H. Statistical validation of image segmentation quality based on a spatial overlap index1: scientific reports. *Acad. Radiol.* **11**, 178–189 (2004).